# RADx Data Hub: A Cloud Platform for FAIR, Harmonized COVID-19 Data

Marcos Martinez-Romero[1,*], Matthew Horridge[1], Nilesh Mistry[2], Aubrie Weyhmiller[3], Jimmy K. Yu[1], Alissa Fujimoto[2], Aria Henry[3], Martin J. O'Connor[1], Ashley Sier[2], Stephanie Suber[3], Mete U. Akdogan[1], Yan Cao[1], Somu Valliappan[2], Joanna O. Mieczkowska[3], Ashok Krishnamurthy[3], Michael A. Keller[2], Mark A. Musen[1] & the RADx Data Hub team

[1] Stanford Center for Biomedical Informatics Research, Stanford University,
Palo Alto, CA 94304, USA
[2] Booz Allen Hamilton Inc., McLean, VA 22102, USA
[3] Renaissance Computing Institute (RENCI), University of North Carolina at Chapel Hill,
Chapel Hill, NC 27517, USA

[*] Corresponding author:

## Abstract

**Background:** The COVID-19 pandemic exposed significant limitations in existing data infrastructure, particularly the lack of systems for rapidly collecting, integrating, and analyzing data to support public health responses. Disparate and non-interoperable data systems led to delays in conducting comprehensive analyses and making timely decisions.

**Objective:** To overcome these challenges, the U.S. National Institutes of Health (NIH) launched the Rapid Acceleration of Diagnostics (RADx) initiative, with the RADx Data Hub serving as a centralized platform for de-identified and curated COVID-19 data. The objective of this paper is to present an overview of the RADx Data Hub's architecture, data management processes, and capabilities.

**Methods:** The RADx Data Hub was developed on a scalable cloud infrastructure designed to comply with the FAIR (Findable, Accessible, Interoperable, Reusable) principles. The Hub integrates various data types, including clinical data, testing results, and social determinants of health, using metadata standards, interoperable formats, and ontology-based tools. It adopts a pipeline for data collection, harmonization, and validation, involving automated and manual quality checks to ensure data integrity.

**Results:** The RADx Data Hub hosts 178 studies across four RADx programs, including contributions from over 100 research organizations across 46 U.S. states and territories. The platform supports data discovery, access, and analysis through tools such as the Study Explorer and Analytics Workbench. These features enable researchers to conduct cross-study analyses and integrate diverse data sources.

**Conclusions:** The RADx Data Hub successfully addresses key data integration challenges by providing a centralized, FAIR-compliant platform for public health research. Its adaptable architecture and data management practices are designed to support secondary analyses and can be repurposed for other scientific disciplines, enhancing preparedness for future health crises.

# Introduction

The COVID-19 pandemic exposed significant shortcomings in the infrastructure required to collect, integrate, and analyze data effectively for public health responses [1,2]. The absence of standardized data and metadata, coupled with the lack of an accessible centralized platform, presented substantial barriers to informed decision-making and interdisciplinary collaboration [3]. In response, the National Institutes of Health (NIH) launched the Rapid Acceleration of Diagnostics (RADx) initiative in 2020 [4], with a key objective to develop such a centralized platform—the RADx Data Hub [5]. This platform aims to aggregate, harmonize, and provide access to data generated by various programs in the overall RADx initiative, enabling data reuse for secondary analyses and facilitating cross-study comparisons. The Data Hub supports innovative research into diagnostic tools, implementation strategies, and health disparities, particularly for underserved populations.

At the core of the RADx Data Hub (hereafter referred to as the Data Hub) is a cloud-based repository grounded in the FAIR (Findable, Accessible, Interoperable, Reusable) guiding principles for data sharing [6]. The machinery and the specifications of the Data Hub are designed to make data (1) findable, through curation, annotation, and support for advanced search and data exploration; (2) accessible, through established privacy and security protocols meant to protect sensitive data while supporting data reuse; (3) interoperable, by leveraging semantic technologies and by defining standard data and metadata models for Data Hub entities; and (4) reusable, by developing routines to validate data quality, techniques to harmonize heterogeneous data, and cloud-based infrastructure to support integrated data analysis on sensitive data.

While the Data Hub was developed specifically to host COVID-19 studies, its underlying technologies were intentionally designed to be repurposed for other domains. The architecture, data and metadata models, software components, and protocols provide a reusable framework that can be adapted to support new systems addressing other scientific research domains. This paper presents the Data Hub as a key resource for secondary analysis of COVID-19 data and details the data management techniques and software tools developed to bring it to fruition as a reliable and accessible data platform.

# Methods

The Data Hub's approach to supporting data reusability and interoperability is built upon four key components: (1) a structured model organizing study data, metadata, and documentation, (2) an end-to-end pipeline for data collection and analysis, (3) a scalable, cloud-based system architecture, and (4) a comprehensive data and metadata harmonization framework. These components are complemented by rigorous data de-identification measures to ensure compliance with established privacy standards, resulting in a flexible foundation that can be adapted for use beyond the COVID-19 context to support a wide range of scientific research domains. The following subsections detail each component and illustrate how they interconnect to form a cohesive, FAIR-compliant infrastructure for scientific data collection, harmonization, and analysis.

## A Model to Organize Study Information

The Data Hub sources its data from four RADx programs: RADx Radical (RADx-rad), RADx Underserved Populations (RADx-UP), RADx Technology (RADx-Tech), and RADx Digital Health Technologies (RADx-DHT) [7]. These programs fund investigators that contribute study information through program-specific Coordination and Data Collection Centers (C)DCCs,[1] each with a distinct mission. RADx-UP, the largest contributor, focuses on health disparities and community-level data collection from underserved populations. RADx-rad explores experimental (radical) technologies for non-traditional COVID-19 detection and surveillance settings. RADx-Tech supports the development and validation of diagnostic technologies. RADx-DHT generates digital health data from wearable devices, mobile apps, and real-time monitoring platforms. Collectively, the Data Hub integrates diverse research efforts, enabling comprehensive analyses through harmonized study data.

The Data Hub stores information for each study, categorized into three overarching components: study data, study metadata, and study documentation (Figure 1). *Study data* constitute the core outputs of each study, capturing the essential variables and results generated during research activities. Study data are organized into one or several file bundles, each comprising three main components: data file, data dictionary, and file metadata:

- *Data file*. A tabular Comma-Separated Values (CSV) document structured according to its corresponding data dictionary. *Variables*, defined in the data dictionary, represent measurable attributes or characteristics (e.g., age, sex, education) and are organized as columns in the data file. Each row corresponds to a data point, such as an experimental sample or study participant, with values recorded for one or more variables. A data dictionary is essential for interpreting the values in the data file.

---

[1] The RADx-UP program includes a coordination function, whereas RADx-rad, RADx-Tech, and RADx-DHT focus solely on data collection. The abbreviation (C)DCCs is used to collectively refer to all of them.

- *Data dictionary*. A CSV document in which each row unambiguously describes the specification of a variable measured or collected by a study or experiment. The columns of the data dictionary provide identifiers for a variable, including properties such as the variable name, label, units of measurement, datatype, and other attributes relevant to specifying a variable's values. The complete specification of data dictionaries used by the Data Hub can be found in the Data Hub Data Dictionary Specification [8]. Data dictionaries are automatically validated against the data dictionary specification upon data submission, and compliance with the specification is strictly enforced. As a result, data dictionaries (and variable metadata, by extension) are amenable to uniform machine processing according to the specification.

- *File metadata*. Each data file is paired with a metadata file that describes its key characteristics, including versioning information, creator details, and summary statistics.

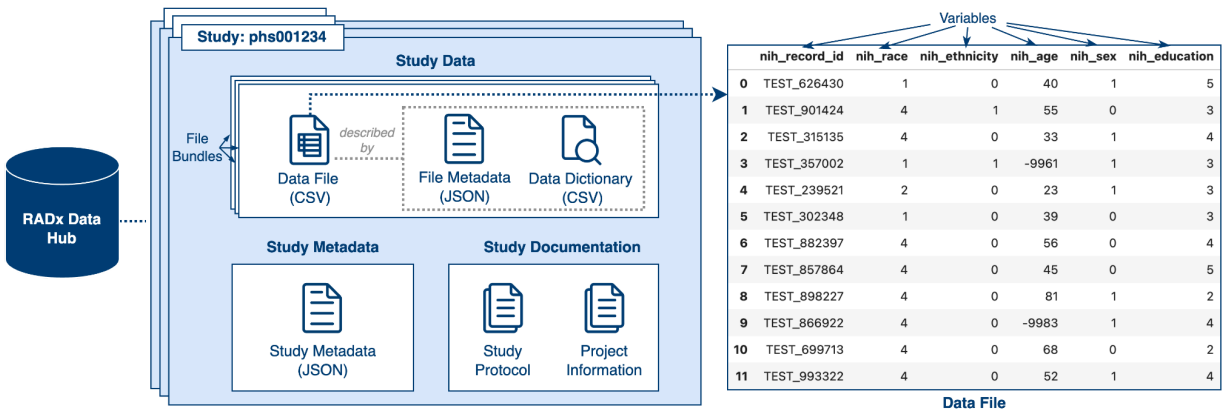

**Figure 1. Components of Data Hub Studies.** This figure illustrates the organization of study information within the Data Hub. Study data (left) are organized into one or more file bundles. Each bundle contains a data file, file metadata, and a data dictionary. Additionally, studies are accompanied by study metadata and documentation. An example data file (right) demonstrates the structure of collected variables, including sample values.

Each study is also paired with a *study metadata* file, offering a comprehensive view of its scope, goals, and content. Study metadata includes details such as the study's geographic and demographic focus, methodologies employed, key outcomes, funding sources, and contributing institutions. By providing a consolidated overview, study metadata facilitates the identification and evaluation of studies that are relevant to specific research questions, while enabling systematic reviews and meta-analyses.

In addition, each study is accompanied by *study documentation* that describes the design and execution processes. The study documentation includes study protocols, project summaries, manual of operations (MOP), and data-collection instruments. By capturing this rich contextual information, the study documentation helps end users understand the study and use the resulting data effectively. Additionally, it offers critical insights into the rationale behind the study design, the measures taken to maintain data quality, and the study's adherence to regulatory and ethical standards, thereby building trust in the data and facilitating responsible data reuse.

## An End-to-End Pipeline for Data Collection and Analysis

Data Hub's end-to-end pipeline (Figure 2) streamlines collection, de-identification, harmonization, validation, curation, storage, and analysis of the data. Data producers, including individual study investigators and research groups, transmit their study data to one of the four (C)DCCs. These (C)DCCs, each aligned with a specific focus within the RADx initiative, play a critical role in the pipeline by standardizing, harmonizing, and preparing the data for inclusion in the Data Hub.

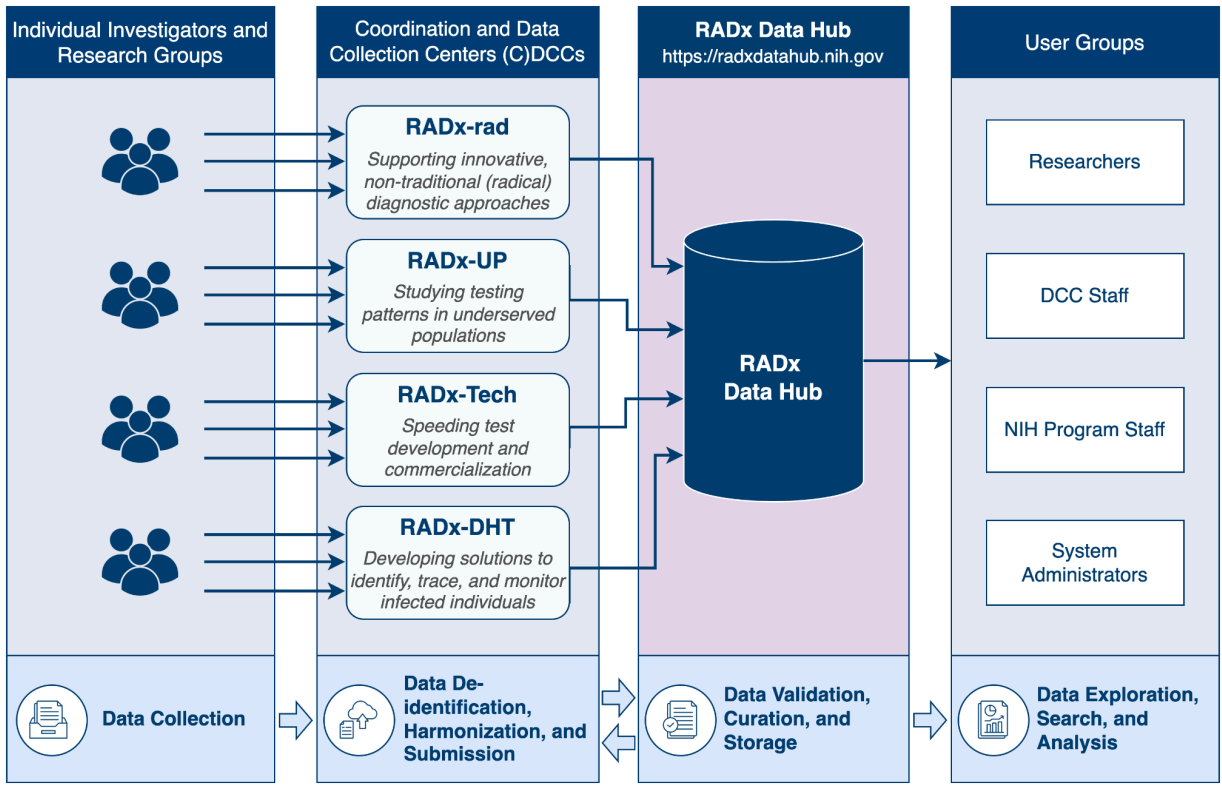

**Figure 2. Overview of Data Hub's end-to-end data pipeline.** Individual investigators and research groups collect and transmit data to the appropriate Coordination and Data Collection Center (C)DCC. At the (C)DCCs, staff de-identify, harmonize, and submit the study data to the Data Hub. Any identified issues are communicated to data contributors, creating a feedback loop for iterative improvements and alignment with Data Hub standards. Once validated, the data are approved, curated, stored in the Data Hub, and then made accessible to users.

After submission to the Data Hub, the data undergo automated validation to detect quality issues, such as non-adherence to standards, incorrect formats, or incomplete metadata. Additionally, the validation process includes automated detection of potential Protected Health Information (PHI) and Personally Identifiable Information (PII) to ensure compliance with privacy and security regulations. These automated steps are followed by manual curation, enabling further refinement of the data. Once these processes are complete, the data are stored and made accessible to users.

## A FAIR Pipeline Using CEDAR and BioPortal

The Data Hub leverages the CEDAR Workbench and the BioPortal ontology repository as foundational tools for metadata management and interoperability.

CEDAR (Center for Expanded Data Annotation and Retrieval) [9] is a suite of web-based tools designed to streamline the creation and management of high-quality metadata. At the core of CEDAR is the concept of metadata templates—structured forms that define the attributes and constraints necessary to describe specific types of data consistently. Templates incorporate controlled vocabularies, specifying permissible values for fields such as demographic characteristics, experimental conditions, and data collection methodologies. Researchers and data contributors use these templates to annotate data with standardized, machine-readable metadata, ensuring adherence to community standards and improving data FAIRness [10].

CEDAR uses the BioPortal ontology repository [11] to integrate ontology terms into metadata templates, ensuring fields are annotated with standardized, semantically rich vocabularies. BioPortal is a comprehensive repository that hosts a vast range of biomedical ontologies, providing standardized terms and definitions across diverse domains to support consistent semantic annotation of data. It provides access to over 1,000 ontologies when building CEDAR templates, enabling users to select standardized terms and definitions to describe study variables. This integration helps ensure that metadata fields, such as "age group" or "diagnostic technology," are annotated with consistent, widely recognized terms, promoting interoperability and reuse of the data across different domains and platforms.

By embedding CEDAR's structured metadata templates and BioPortal's rich vocabulary resources into its metadata workflow, the Data Hub provides a comprehensive framework for FAIR compliance. The following subsections detail how this framework was used to help ensure the findability, accessibility, interoperability, and reusability of RADx data.

## Findability: Exploring and Searching Study Data

One of the primary purposes of the Data Hub is to provide a platform for researchers to find study data appropriate for their secondary analyses. While access to study data is governed by the NIH database of Genotypes and Phenotypes (dbGaP) [12] and requires formal approval, the Study Explorer component of the Data Hub (Figure 3) is openly available. It allows users to browse the list of studies, explore their characteristics, and assess their suitability before submitting a data access request through dbGaP. The Study Explorer offers a range of search functionalities, including free-text search, conditional queries using multiple criteria, and options to sort, filter, and download results. Users can filter results using various facets, including study domain, sample size, and data collection method. Studies can also be located directly by querying for their title or other study attributes.

**Figure 3. Study Explorer.** The figure shows a screenshot of the Data Hub Study Explorer, with the "Study Population Focus" filter applied to retrieve 80 studies on underserved or vulnerable populations. A search box at the top, equipped with an autocomplete feature, allows users to quickly find studies by entering relevant keywords. The filtered results are displayed in a table, which includes information such as study names, dbGaP accessions, estimated cohort sizes, NIH institutes or centers sponsoring the study, and the RADx Data Program responsible for the study.

Upon locating a study of interest, users can view its details on the Study Overview page (Figure 4). This webpage provides a concise summary of the study, offering access to the key components: study metadata, documentation, and data files, including file metadata and data dictionaries. Study metadata can be downloaded in machine-readable formats, such as JSON and YAML, for easy integration into research workflows. To simplify metadata viewing, the Data Hub provides an interactive Metadata Viewer (Figure 5) powered by CEDAR's Embeddable Editor [13], which includes links to ontology terms from the BioPortal ontology repository [11,14]. This tool allows users to inspect all metadata associated with a study and to verify that the study aligns with their research goals. Additionally, the Study Overview page displays the variables used in each file, enabling researchers to identify variables relevant to their analyses or to search for related studies that share the same variables. Furthermore, each study on the Data Hub is assigned a unique DataCite Digital Object Identifier (DOI), a persistent string that provides a permanent reference, ensuring the study can be reliably located and accessed within the Data Hub.

**Figure 4. Study Overview Page.** This figure provides an overview of the information available for a selected study within the Data Hub. The panel at left summarizes key metadata, such as the dbGaP study accession, RADx Data Program, study domain, and data collection methods. The panel at right is divided into "Study Documents" and "Data Files." The "Study Documents" section lists supporting files, such as study protocols and README files. The "Data Files" section contains the data files and their corresponding metadata, along with a button to request access to the data through the study's dbGaP page.

**Figure 5. Metadata Viewer.** The left side of the figure displays metadata for a specific data file, with terms constrained to ontology terms, linked to BioPortal for further exploration. The right side demonstrates a link from the "subject identifier" field to the corresponding term in the Medical Subject Headings (MeSH). Metadata are organized into expandable and collapsible sections for easy navigation, with "help" text providing guidance and context. The Metadata Viewer renders a CEDAR template designed for Data Hub files, ensuring consistency and accessibility across metadata records.

**Accessibility: Ensuring Data Availability and Secure Access**

Protecting sensitive health-related data requires robust control mechanisms to ensure compliance with ethical standards, government regulations, and data use agreements. With the exception of RADx-DHT, whose data reside in the RAPIDS repository [15], all study data are stored in the Data Hub. Access to files from the Data Hub is reviewed and authorized by the NIH Data Access Committee (DAC) with an established Data Use Agreement through dbGaP [12]. This agreement outlines specific terms for data usage, including requirements for privacy protections and ethical conduct. Additionally, users must follow the Data Hub User Code of Conduct [16], which sets standards for responsible data handling.

To formally request access to study-level data, users must log into the Data Hub using their electronic Research Administration (eRA) Commons or NIH credentials, locate the desired study, and submit an access request. This requirement ensures that only verified researchers and authorized personnel can access controlled datasets, in compliance with NIH security and data governance policies. While this login process may pose a barrier for new users, it helps protect sensitive research data and aligns with NIH's data access standards. Any individual with Principal Investigator (PI) status can obtain an eRA Commons login and dbGaP authorization. Once access is approved by the DAC, PIs can manage permissions for their team members via dbGaP. Tutorials on the data request process are available on the Data Hub Resource Center webpage [17] to assist users with each step.

Study metadata, summary statistics, and documentation remain openly accessible through the Study Explorer and Study Overview pages. These resources allow users to evaluate the relevance of study data and variables before submitting a formal data access request, ensuring they can determine whether the available datasets align with their research needs without requiring authentication.

**Interoperability: Enforcing Data and Metadata Standards**

The Data Hub establishes a foundation for interoperability by adopting globally recognized data formats and metadata standards, alongside a standardized set of NIH Common Data Elements (CDEs). CDEs are defined as combinations of standardized questions (variables) paired with an enumerated set of possible responses (values) that are "common" across multiple studies [18]. Variables in data files are mapped to these CDEs to ensure consistent (i.e., harmonized) interpretation across studies.

The CDEs used in the Data Hub are defined in the Data Hub Global Codebook [19]. This listing includes 133 core CDEs that provide a consistent framework for representing essential variables, such as demographic and participant health information, across all RADx studies. These CDEs are organized into 12 categories, including Race, Ethnicity, Age, Sex, Education, Domicile, Employment, Insurance Status, Disability Status, Medical History, Symptoms, and Health Status. The Data Hub team is responsible for performing harmonization to these CDEs where possible by employing a harmonization methodology detailed in the "Data and Metadata Harmonization" section. Submitted studies, however, are not limited by these CDEs and may include additional variables collected by the original researchers.

Beyond the provision of CDEs, the Data Hub employs standardized file formats and metadata specifications to further enhance interoperability. Data files are provided in CSV format, organized in a consistent structure where variables are represented as columns and rows as individual data points. Data dictionaries follow the RADx Data Dictionary Specification [8], also in CSV format, ensuring uniform interpretation of variables. Metadata for data files and studies are formatted in JSON-LD, adhering to the CEDAR model specification [20]. Additionally, study documentation is provided in standard formats, such as PDF and plain text, ensuring accessibility and compatibility for a wide range of users.

In practice, the (C)DCCs collaborate closely with the Data Hub team to ensure compliance with these global standards. The (C)DCCs conduct the initial harmonization, while the Data Hub team maintains the Global Codebook and performs additional post hoc harmonization as needed before studies are made publicly available. These efforts aim to enhance the interoperability of study data within the RADx ecosystem and with external data repositories and analytical platforms. This comprehensive standardization approach maximizes the data's utility for research and public health applications.

**Reusability: Prioritizing Data Quality and Enabling Secondary Data Analysis**

The Data Hub promotes study data reusability by focusing on two critical aspects: ensuring high data quality and providing an integrated platform for secondary data analysis.

Data Quality

To ensure study data utility and reliability, the Data Hub employs a rigorous quality assurance and quality control (QA/QC) process. This process begins with close collaboration between the Data Hub team and (C)DCC staff during data submission, ensuring adherence to the data and metadata standards for metadata files, data dictionaries, and documentation described in previous sections. QA/QC reviews are performed rigorously before data and metadata are made publicly findable, ensuring compliance with privacy policies, alignment with the Data Hub's standards, and readiness for harmonization and secondary use.

A key aspect of this process involves evaluating file bundle completeness, identifying and addressing missing values, and standardizing metadata formats to ensure consistency across data files and studies. While the quality of the data reflects the efforts of the original data producers, the Data Hub enhances value by applying harmonization practices during ingestion and by documenting any data and metadata gaps or limitations. These enhancements improve transparency, facilitate cross-study comparisons, and make the data more reliable for downstream analyses.

Data Analytics Workbench

The Data Hub features a cloud-enabled platform called the Analytics Workbench as a key feature for promoting data reuse (Figure 6). This workbench is a versatile platform, powered by AWS SageMaker Studio technology, that enables users to download or transfer data, to create personalized workspaces, and to deploy advanced analytical tools. By leveraging this platform, researchers can securely access, manipulate, and analyze RADx data.

The workbench provides users standard data analysis tools such as RStudio, Python, and JupyterLab, all readily accessible without additional setup. Additionally, users can extend the platform's capabilities by accessing add-ons such as Data Wrangler, a "no-code" solution for data transformation, analysis, and visualization, and SAS Viya Analytics Pro, a comprehensive cloud-hosted statistical environment for data access, transformation, analysis, visualization, and mapping. The workbench also supports integration with external platforms such as DockerHub and GitHub, enabling users to tailor their workflows to specific research needs.

Users are able to view the Data Hub's Public Data page and Analytics Workbench. The Public Data page provides access to data files that are openly available, including files containing synthetic data. Users can download these data files directly or transfer them to the Analytics Workbench for further analysis. This feature allows researchers to explore the platform's tools and capabilities before requesting access to protected study data. The Analytics Workbench is shown on the right-hand side of Figure 6, where a Jupyter Notebook environment allows analysis of the distribution of the subject's education levels within a synthetic data file. This example highlights the Workbench's capacity to facilitate secure data analysis and visualization, offering a flexible environment equipped with various analytical tools.

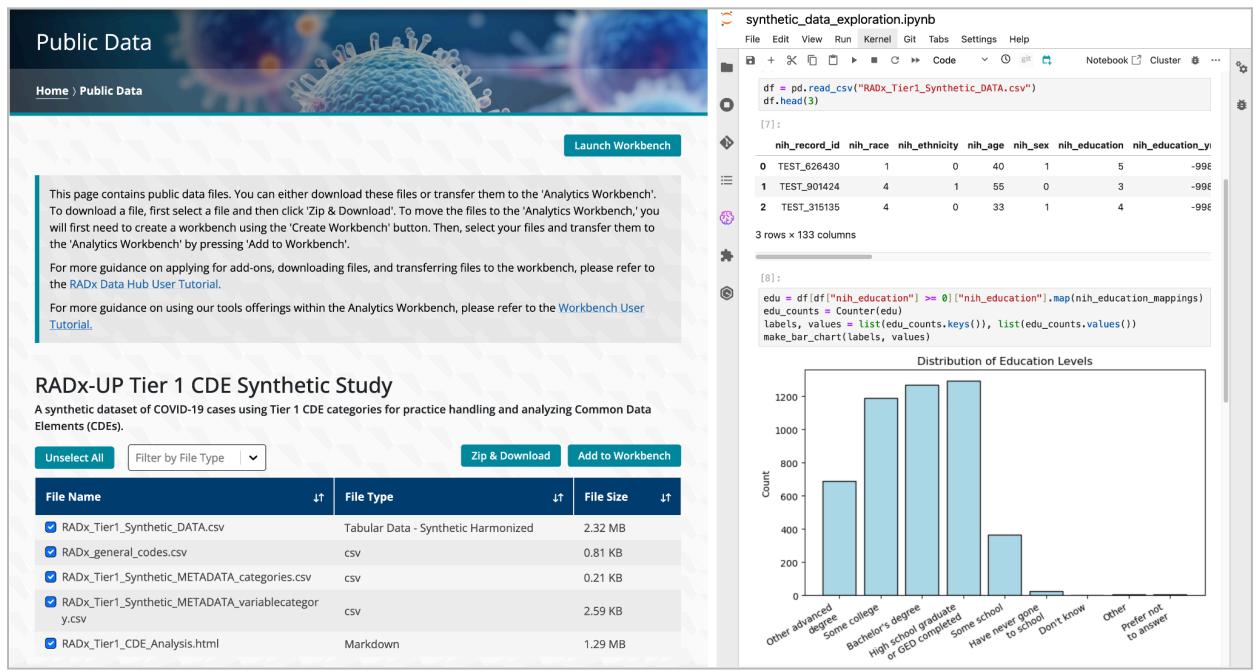

**Figure 6. Public Data and Analytics Workbench.** The left side highlights the Public Data Page, where users can access and transfer public data files to the Workbench. The right side displays the Analytics

Workbench using a Jupyter Notebook to analyze the distribution of education levels in synthetic data, showcasing the platform's data analysis and visualization capabilities.

## System Architecture

The Data Hub is a cloud-based platform designed to provide a scalable, secure, and FAIR-compliant infrastructure for collecting, harmonizing, and analyzing COVID-19 data. The Data Hub architecture supports the flow of diverse data types from data producers to a centralized data repository, enabling researchers and stakeholders to leverage data effectively. The architecture comprises two functional layers—the frontend and backend—both hosted on AWS cloud infrastructure and interconnected through secure APIs (Figure 7). This architecture is designed to scale with increasing data and user demands, while maintaining high performance and availability.

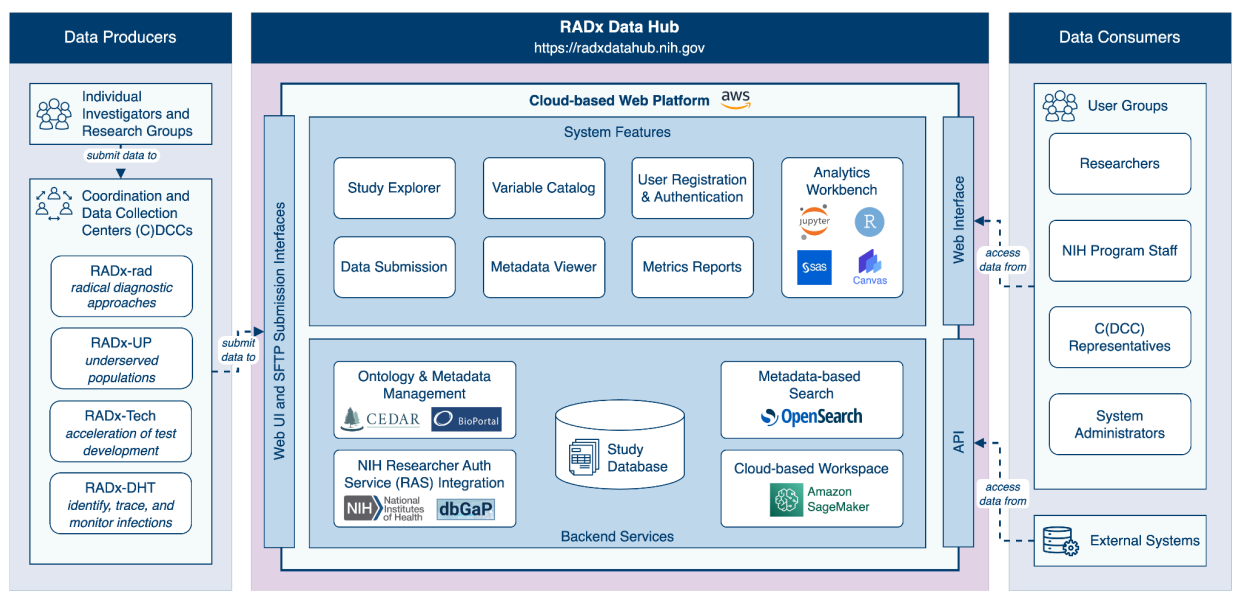

**Figure 7. Overview of the Data Hub's software architecture.** Data producers submit study data to the Data Hub, hosted on AWS. The Data Hub features backend services for ontology and metadata management, metadata-based search, and secure storage, ensuring data is harmonized, validated, and curated. Its main features include study and variable search, exploration, and advanced analytics, while an integrated workbench provides researchers with a scalable environment for conducting advanced analyses and running machine learning workflows.

The frontend layer provides user-facing functionalities through a modern web interface built with React and Next.js. The frontend layer supports dynamic page rendering, device compatibility, and accessibility requirements (including Section 508 compliance [21]). Key features include the Study Explorer, Variable Catalog, and Metadata Viewer, which allow users to browse, search, and explore data and metadata, and access secure user registration and authentication systems. Additional features, such as the Analytics Workbench, integrate Jupyter, R, and SAS to facilitate advanced data analysis and visualization. Metrics Reports further enhance this layer by providing actionable insights into data usage, quality, and platform performance, supporting effective system monitoring.

The backend layer is implemented using a microservice architecture powered by Spring Boot and Python. It handles business logic and workflows essential for data ingestion, harmonization, and search. Core components include (1) Submission Services for validating and processing incoming data, (2) Search Services powered by AWS OpenSearch for efficient metadata-based retrieval, and (3) an integrated Analytics Workbench leveraging AWS SageMaker for scalable machine learning workflows. Security and compliance are maintained through robust identity management systems, including NIH Researcher Auth Service (RAS) integration and role-based access control (RBAC).

The system's software architecture is designed for flexibility and scalability to accommodate additional COVID-19 study data, facilitate the addition of new features, and support the potential expansion into other health domains. A Postgres relational database, managed via the AWS Relational Database Service (RDS), stores metadata, while raw data files (without metadata) are housed in Amazon S3. This separation ensures efficient querying and modular scalability. AWS OpenSearch indexes metadata to provide fast and precise search capabilities. CEDAR and BioPortal enable metadata annotation and semantic interoperability, ensuring adherence to FAIR principles.

The cloud infrastructure is built on AWS, leveraging serverless technologies, such as AWS Fargate and Lambda for containerized microservices and tasks and Elastic Load Balancers for high availability. Development, testing, and production environments are isolated to enhance security and streamline deployment workflows. Automated CI/CD pipelines enable rapid and reliable updates to the system, maintaining operational stability and performance.

## Data and Metadata Harmonization

Harmonization is a key component of the Data Hub's mission to standardize study data, enabling cross-study analyses and the re-exploration of data within individual studies. Harmonization can be approached in two ways: (1) *prospective harmonization*, where common data and metadata standards are defined and established ahead of data collection, and (2) *retrospective harmonization*, which enforces common data standards and ensures consistent representation of semantically equivalent data through careful curation and processing after the data have been collected [22–24]. Given the urgent, time-sensitive nature of the RADx initiative, prospective harmonization was not feasible. Developing and implementing advanced standards prior to data collection would have delayed critical public health research efforts. Therefore, retrospective harmonization was adopted as a pragmatic solution, enabling the standardization of diverse information while ensuring timely data availability for the broader goals of the RADx initiative.

The harmonization process within the Data Hub, as illustrated in Figure 8, involves two main steps: (1) mapping data dictionaries to the Data Hub's Global Codebook [19] and (2) transforming data and metadata to align with standardized templates.

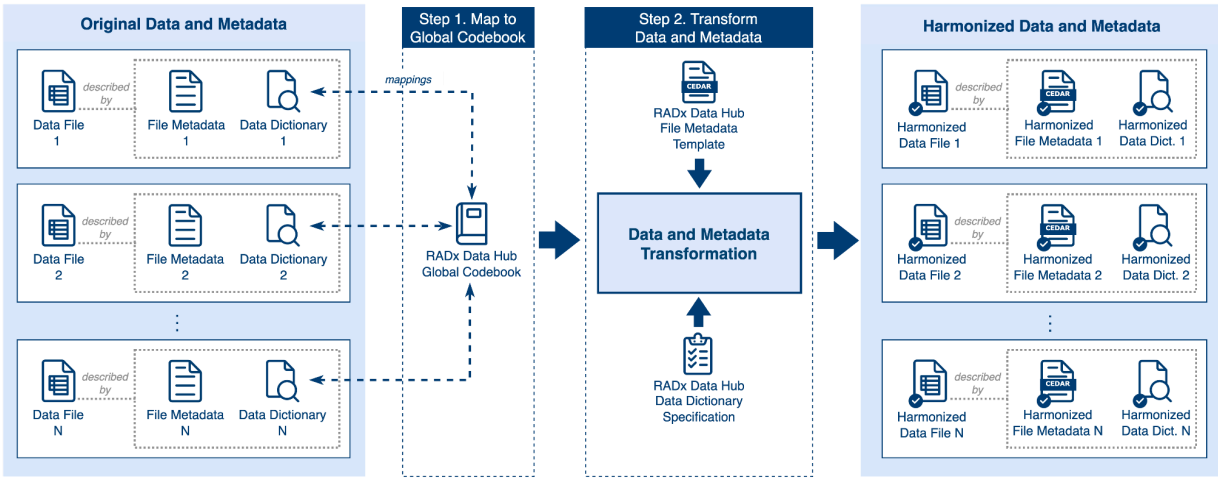

**Figure 8. Harmonization Workflow.** Harmonization of data files and metadata files within the Data Hub involves a two-step process. The Data Hub receives original data files, accompanied by their respective metadata files and data dictionaries (at left). In Step 1, the original data dictionaries are mapped to the Data Hub Global Codebook for CDEs, standardizing variable definitions and allowable values across studies. In Step 2, the original data and metadata are transformed into harmonized formats: data files are transformed using mappings from Step 1; metadata files are aligned to a CEDAR metadata template; and data dictionaries are converted to the Data Hub Data Dictionary Specification (center). This process ensures the resulting harmonized data files, metadata, and dictionaries are consistent and interoperable across the Data Hub (at right).

The harmonization workflow begins with mapping original data dictionaries provided by investigators to the Global Codebook of RADx Common Data Elements (CDEs), which serves as a unified framework for standardizing variable definitions and values across studies. The identification and creation of these mappings are performed manually by the Data Hub team, leveraging their expertise to account for the variability and complexity of submitted data. The resulting mappings are stored in the Global Codebook, creating a centralized, reproducible resource that supports ongoing and future harmonization efforts.

Following the mapping process, data and metadata are transformed to achieve harmonization, ensuring consistency and interoperability across datasets. Data files are transformed using the mappings identified previously. Study-level metadata (e.g., Principal Investigator information, funding sources, institutional affiliations) are aligned with a dedicated CEDAR metadata template [25]. Similarly, data file metadata (e.g., file name, version, creator) are structured using a CEDAR template [26] inspired by the DataCite Metadata Schema [27]. The DataCite Metadata Schema provides a widely recognized list of standard core metadata properties for citation and retrieval purposes, which the CEDAR template augments by incorporating additional fields tailored to specific needs, such as detailed funding information and file characteristics. By leveraging the strengths of the DataCite standard while extending its functionality, the CEDAR template ensures that metadata are both globally compatible and capable of meeting diverse research requirements. Additionally, data dictionaries are converted to the Data Hub Data Dictionary Specification [8], ensuring uniformity in structure and representation.

Figure 9 illustrates an example of data harmonization for the variable *edu_years_of_school*, found in an original RADx-UP data file, representing a study subject's level of education. On the left, the original variable includes values and corresponding definitions outlined in the RADx-UP data dictionary. For example, a value of "3" represents "9th to 12th grade, no diploma," whereas "4" corresponds to "High school graduate or GED completed."

The center panel demonstrates how these original values and definitions are mapped to the corresponding CDE, *nih_education*, within the Data Hub Global Codebook. For instance, the value "4" in the original data file maps to "2: High school graduate or GED completed" in the Global Codebook, ensuring semantic consistency.

Finally, the right panel shows the harmonized data file, where the variable has been renamed *nih_education* and the values updated according to the mappings. This harmonized format ensures that the data is consistent and interoperable within the Data Hub, facilitating cross-study analysis and integration.

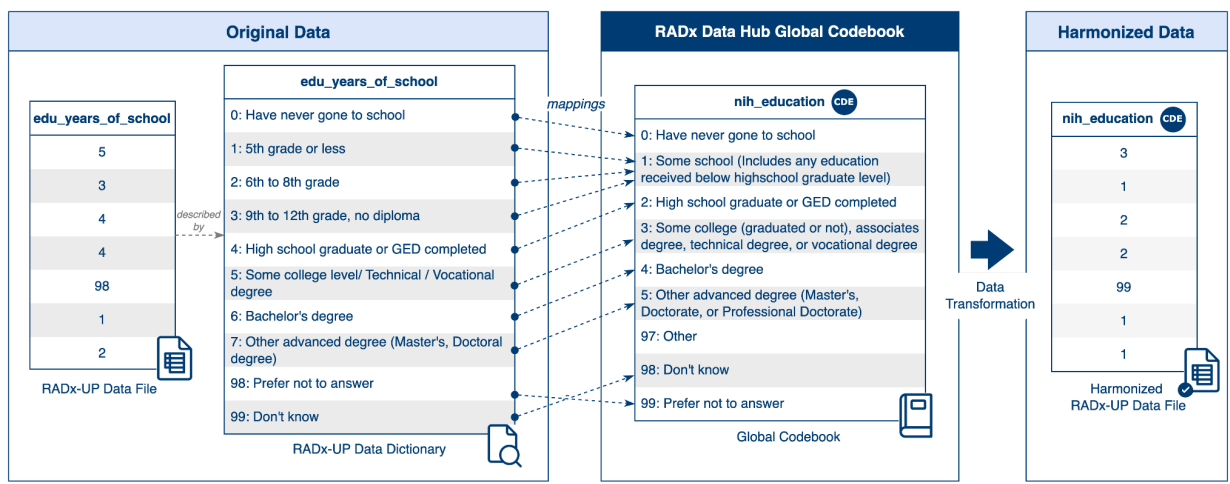

**Figure 9. Data Harmonization Example.** The variable *edu_years_of_school*, representing an individual's level of education in a RADx-UP study, can be mapped to the Data Hub Global Codebook of CDEs. The original values and definitions, outlined in the RADx-UP data dictionary (at left), correspond to the CDE *nih_education* in the Data Hub Global Codebook (center). The harmonized data file (at right) reflects the updated variable name and standardized values.

The harmonization process is supported by rigorous QA measures conducted by the Data Hub team. Automated tools and structured workflows check for compliance with CDE standards, formatting specifications, and metadata guidelines. Any discrepancies or issues are communicated back to data contributors, creating a feedback loop that promotes iterative improvements and alignment with the Data Hub's standards (Figure 2). This QA process ensures harmonized data are reliable, consistent, and ready for reuse.

The Data Hub stores both original and harmonized versions of data files. Harmonized data are recommended for cross-study analyses and data re-exploration due to their standardized format and semantic consistency. However, the original raw data are also retained and made accessible for users requiring unmodified data for specialized research needs. This dual offering supports diverse research applications while ensuring the integrity and reusability of the Data Hub's data resources.

**Data De-Identification**

To protect participant privacy, the (C)DCCs apply a comprehensive de-identification process [28] to study data before their submission to the Data Hub. This process ensures data are anonymized while preserving their scientific utility for secondary research. The (C)DCCs are tasked with balancing PHI and PII removal with the retention of sufficient data to maintain their research value. The Data Hub leverages Amazon Macie at the time of data submission to automatically detect and flag potential PHI/PII. Amazon Macie uses machine learning and pattern matching to identify sensitive data, allowing (C)DCCs to review and apply additional de-identification steps as needed before the data are ingested into the system. This automated screening process helps ensure that all data submitted to the Hub meet privacy and regulatory requirements.

The de-identification process follows a standard operating procedure that incorporates several key techniques. Direct identifiers, such as names and addresses, are redacted entirely to eliminate any risk of re-identification. In compliance with HIPAA rules, ZIP codes are generalized by retaining only the first three digits, unless the area's population is 20,000 or fewer, in which case the ZIP code is replaced with "000." Dates are shifted by a consistent interval to obscure precise temporal details while preserving temporal relationships within the data. Sites within studies are anonymized using random codes, and participant ages are altered: individuals under one year old are recorded as "0," ages 21 to 89 are modified by adding or subtracting 2 years, and ages 90 or older are top-coded as "90."

These procedures are meticulously documented by the (C)DCCs to ensure traceability, replicability, and accountability. By applying these de-identification standards, along with automated PHI/PII detection at submission time, the Data Hub safeguards participant privacy while enabling meaningful secondary use of the data for public health research.

# Results

As of January 2025, the Data Hub hosts 178 studies and over 1,100 data files contributed by researchers from more than 100 organizations across 46 U.S. states and territories. The Data Hub represents one of the largest coordinated efforts to centralize COVID-19-related public health data, supporting both immediate and long-term responses to health crises. By adhering to FAIR principles, the Data Hub enhances access to high-quality, standardized study data for secondary analyses and public health research.

The studies stored in the Data Hub cover a broad range of domains, including community-based research, diagnostic technologies, and surveillance in underserved populations. Metadata and documentation for these studies are openly accessible through the Study Explorer, enabling researchers to evaluate the relevance of datasets prior to requesting full access via dbGaP [12]. This governance ensures compliance with privacy protections and ethical standards while facilitating secure access to sensitive health data.

Figures 10 to 13 show the distribution of studies by domain, population focus, data collection method, and study design. Stacked bar charts highlight the contributions from the different RADx programs, illustrating the diverse focus areas and methodological approaches across RADx initiatives. The RADx-UP program, with its emphasis on health disparities, contributes a significant portion of these studies. Many studies focus on underserved populations, including racial minorities, older adults, and individuals with low socioeconomic status. Data collection methods, such as surveys, wearable device monitoring, and diagnostic testing, offer opportunities for cross-study integration. The majority of studies follow observational designs, with longitudinal cohort studies providing insights into long-term health outcomes.

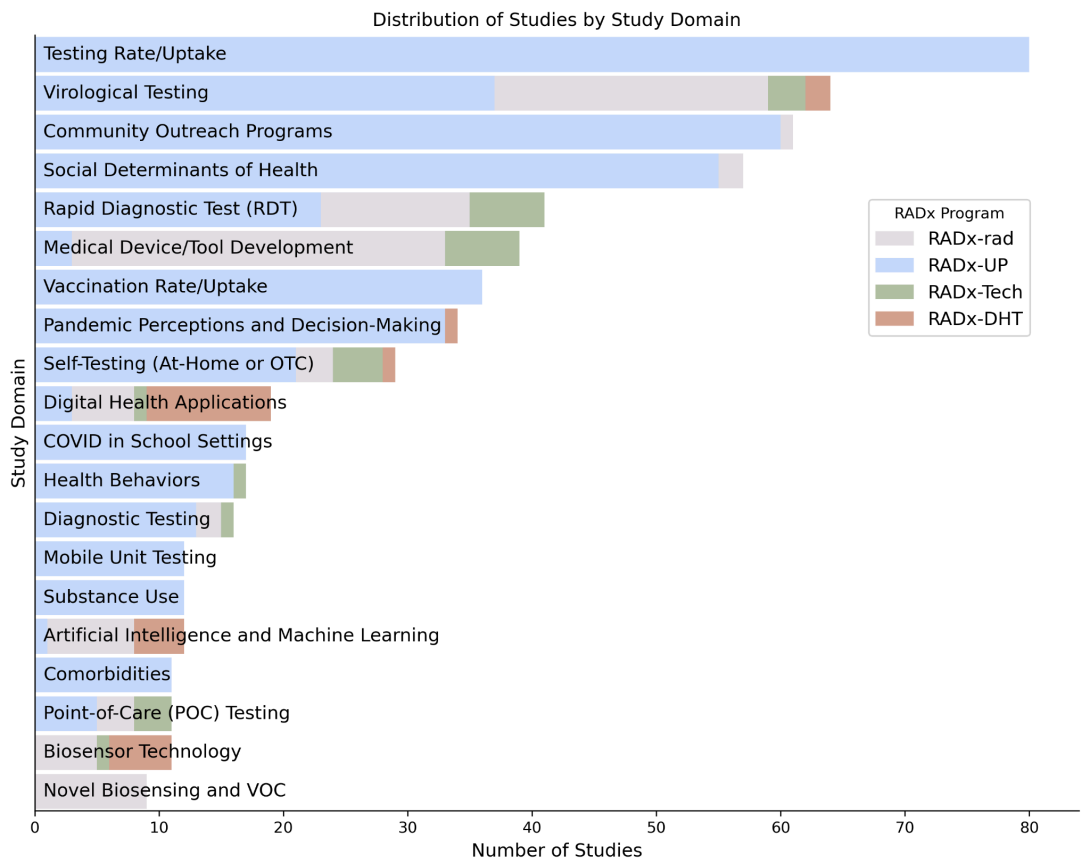

**Figure 10. Distribution of Data Hub studies by study domain.** This histogram illustrates the distribution of studies across various research domains, highlighting the diverse focus areas of RADx studies. RADx-UP has the highest representation, emphasizing its focus on community-based health research. The substantial presence of RADx-rad studies indicates strong interest in non-traditional COVID-19 detection technologies.

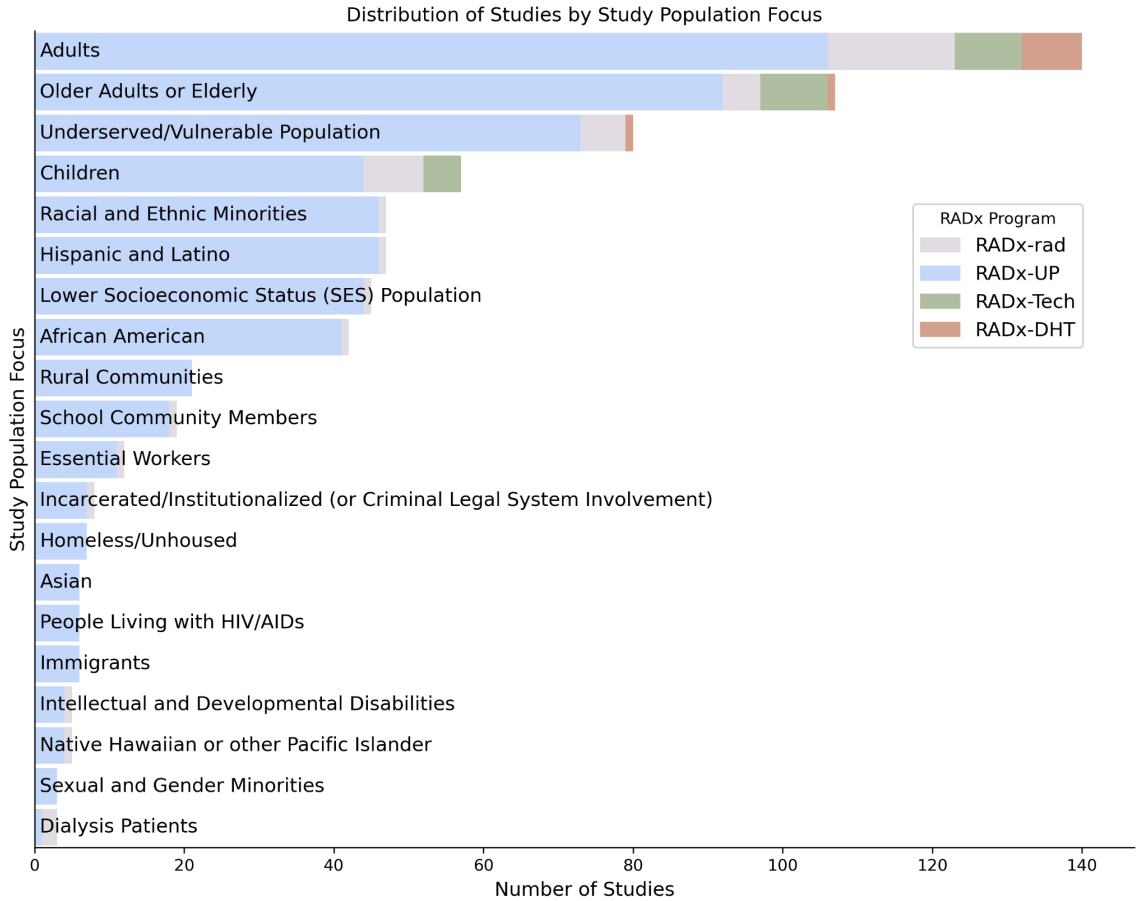

**Figure 11. Distribution of Data Hub studies by population focus.** This figure presents the breakdown of studies based on population focus. The histogram shows that a significant proportion of studies target underserved and disproportionately affected populations, reflecting the primary goal of RADx-UP. The presence of studies focused on general populations and specific at-risk groups, such as older adults and children, highlights the breadth of data available for comparative analysis.

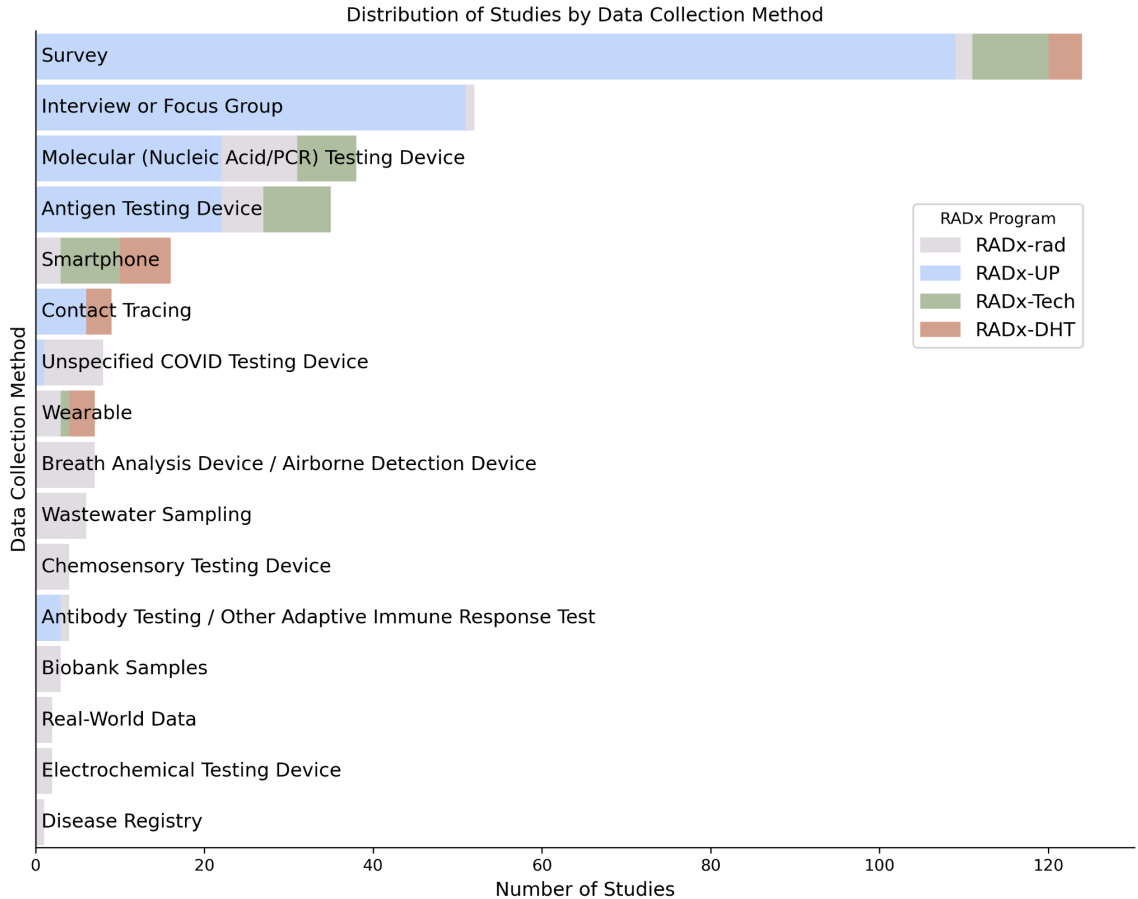

**Figure 12. Distribution of Data Hub studies by data collection method.** This histogram categorizes studies based on their data collection approaches. The visualization highlights the prevalence of survey and interview-based data collection, particularly within RADx-UP studies. The diversity of data collection methods suggests opportunities for cross-study analyses, integrating multiple data sources for a more comprehensive understanding of pandemic-related health outcomes.

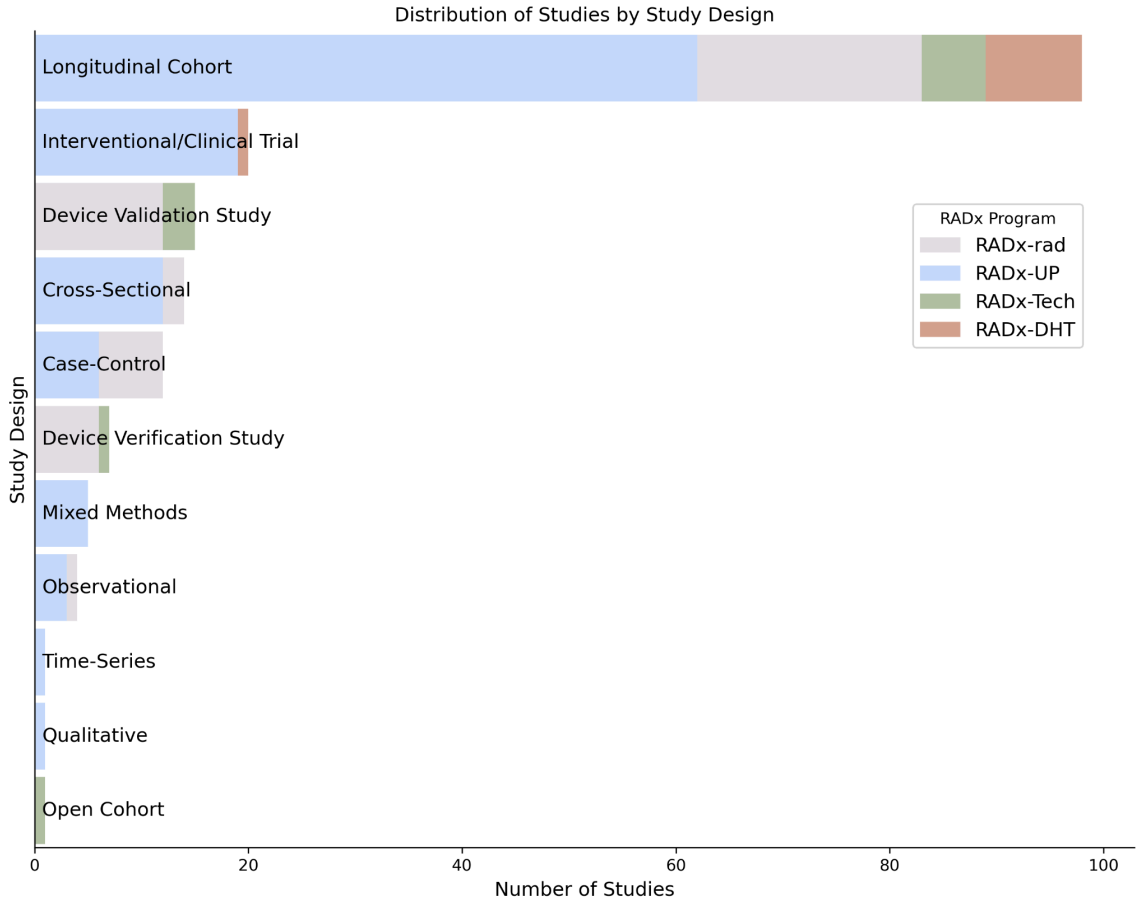

**Figure 13. Distribution of Data Hub studies by study design.** This figure illustrates the variety of study designs employed across Data Hub studies. The histogram reveals that most studies follow an observational design, with longitudinal cohort studies being the most common.

The Data Hub streamlines data discovery through tools like the Study Explorer, Variable Catalog, and Metadata Viewer. Each study is assigned a persistent Digital Object Identifier (DOI) to support long-term access and citation. The platform's metadata-driven infrastructure enables researchers to efficiently search for and explore studies and variables, reducing barriers to data reuse and collaboration.

The Analytics Workbench offers a secure, cloud-based environment where researchers can access harmonized datasets and perform analyses using tools like Jupyter and RStudio. By enabling data manipulation, visualization, and advanced analyses directly on the platform, the Workbench reduces local resource demands and improves reproducibility. Researchers can analyze trends, compare study outcomes, and conduct large-scale investigations efficiently without managing their own infrastructure.

The Data Hub's ability to centralize and harmonize data is crucial in addressing key public health questions related to the COVID-19 pandemic. By enabling multi-study analyses, the Data Hub advances research on health disparities, diagnostic effectiveness, and public health response strategies, reinforcing its role as a crucial resource for pandemic preparedness and response, with significant potential for long-term adaptation to future public health challenges.

## Discussion

The Data Hub provides a centralized platform to support rapid public health response to the COVID-19 pandemic, overcoming the challenges posed by fragmented public health data systems. By promoting data sharing, reuse, and analysis, the Data Hub enhances data usability and fosters collaboration among researchers, public health officials, and policymakers, ultimately strengthening the ability to respond swiftly and effectively to public health crises such as the COVID-19 pandemic. The Data Hub employs modern integration techniques and robust data governance practices to ensure data security, privacy protection, and compliance with regulatory standards. The infrastructure and data management techniques developed as part of the Data Hub serve as invaluable tools to combat future public health events by maximizing the utility and impact of shared data within the public health and research communities.

The Data Hub aims to enhance the findability and accessibility of data through rich metadata, allowing researchers to quickly assess whether the collected data aligns with their research interests and topic areas. Moreover, the curation and standardization of data throughout the platform is intended to ensure data quality, to resolve inconsistencies and heterogeneity where possible, and to identify missing data to provide cleaner, ready-to-use data. Standardizing data formats and metadata enhances interoperability and facilitates data integration, enabling the creation of aggregate data files for large-scale cross-study analyses. These capabilities are vital for evaluating and understanding the COVID-19 pandemic response, for addressing disparities in intervention outcomes, for advancing biomedical technologies such as testing and rapid detection, and for preparing for future pandemics [29–31].

### Limitations and Lessons Learned

While the Data Hub has made significant strides in facilitating data sharing and reuse, its development and implementation have highlighted key challenges in managing a large-scale, multi-source data platform. These experiences provide valuable insights into the inherent limitations that arise from the Data Hub's scope and operational framework.

*Data Hub standards evolve over time through ongoing harmonization and expert curation.* The standards for data and metadata within the Data Hub continuously evolve to improve interoperability, usability, and compliance with FAIR principles. Despite the platform's relatively short existence, significant efforts have been required to standardize and harmonize data and metadata after their initial ingestion. The diversity of RADx data, which range from population health surveys to wastewater sampling, presents challenges in maintaining consistency across heterogeneous sources. Data stewards play a critical role in guiding researchers through the submission process, ensuring that study contributions align with established metadata models and CDEs. Standardization efforts extend beyond initial ingestion, with retrospective harmonization processes mapping study variables to the Data Hub Global Codebook. Post hoc curation and validation remain ongoing as the Data Hub team continuously refines data quality, corrects inconsistencies, and integrates updated standards to support scalable, reusable research across disciplines.

*Effective collaboration and coordination with data contributors is essential.* The quality of data hosted by the Data Hub is inherently influenced by the variability in resources and priorities among data contributors. While researchers diligently fulfill their primary research objectives and meet funding requirements, ensuring comprehensive adherence to the Data Hub's standards can sometimes be secondary. Consequently, some data contributions may require additional curation or, in some cases, may be accepted as-is to ensure completeness of the repository. These dynamics underscore the importance of ongoing dialogue and collaboration with (C)DCCs to refine processes and enhance data quality standards.

*Facilitating access to data is essential for maximizing usability.* Accessing Data Hub studies involves adherence to dbGaP's data use agreement, adherence to the Data Hub's user code of conduct, and approval from dbGaP's data access committee. While these safeguards are critical for ensuring compliance with ethical and legal standards, the need to go through a separate system and meet administrative requirements can pose challenges for researchers who must evaluate the possible utility of data based only on study overviews, metadata, or associated publications before initiating access requests. Streamlining and optimizing the data access process is a key area of focus to reduce barriers and support researchers in efficiently leveraging the Data Hub's rich resources for secondary analysis.

*Post hoc harmonization presents unique challenges.* Harmonization efforts so far have focused on human participant studies, resulting in a core set of CDEs suited to such research. However, the Data Hub contains a wide array of data, spanning from human health to environmental sampling, with data that hold potential applications beyond those initially anticipated. As a result, data sets within the Data Hub that cover research domains beyond the scope of the core CDEs, such as wastewater studies or tracking technologies, may benefit from additional post hoc harmonization by the user.

## Conclusions and Future Work

The Data Hub addresses critical challenges in public health data management by providing a centralized, FAIR-compliant platform that integrates diverse datasets for secondary analyses. Through its scalable architecture, metadata-driven workflows, and robust privacy measures, the Data Hub has become a pivotal resource for facilitating cross-study research. By enabling rapid access to harmonized COVID-19 data, the platform advances research efforts on health disparities, diagnostic tools, and pandemic preparedness. The Data Hub will continue to evolve as an essential resource for public health research, with ongoing efforts aimed at improving data and metadata quality, usability, and accessibility.

The Data Hub is continuously evolving to meet the dynamic needs of public health research. Efforts to improve data and metadata quality, usability, and accessibility are ongoing. A notable initiative is the development of a harmonization software framework to streamline the integration of heterogeneous data. Currently, the Hub relies on a manually curated set of mappings between (C)DCC variables and CDEs in the Global Codebook. The proposed framework will automate the mapping of variables to CDEs and support the harmonization of custom variables, significantly expanding the versatility and usability of the Data Hub.

Improving data findability remains a top priority. At present, the Variable Catalog operates separately from the Study Explorer, complicating the search for specific variables and related studies. We are developing a single search user interface that will unify study and variable searches, enabling users to identify both studies and the specific variables they contain. This interface will be enhanced by incorporating Dug [32], an innovative semantic search technology from the BioData Catalyst ecosystem. Dug utilizes natural language processing (NLP) and ontological knowledge graphs to provide context-aware search results. For example, a search for "Long COVID symptoms" could identify studies tagged with the synonym "post-acute sequelae of SARS-CoV-2 infection," significantly improving the accuracy and breadth of search outcomes.

The Analytics Workbench is another key feature, providing a secure cloud-based environment for data analysis. While users can download data for local use, the Workbench offers advantages such as enhanced security, controlled access, and integrated analytical tools like Jupyter Notebook, RStudio, and SAS Viya Analytics Pro. From a cost-benefit perspective, the Workbench minimizes redundant data downloads and streamlines workflows, saving time and computational overhead. However, cloud infrastructure requires ongoing investment, and the Data Hub team is exploring models to ensure sustainability while maximizing its benefits for researchers.

Additional enhancements in development include a CEDAR-based data dictionary viewer and enhanced visualization tools through BioPortal, allowing for more intuitive exploration of RADx terms and their associations. These improvements will be complemented by external authority validations for metadata, such as ORCID [33] for researcher identifiers and RoR[34] for organizational affiliations.

While initially developed to address the COVID-19 pandemic, the Data Hub's infrastructure is designed to be adaptable to other public health challenges. The underlying architecture, data and metadata models, software components, and protocols can be repurposed to support other research initiatives, as we intend for the Data Hub to be a long-term resource for the broader scientific community.

## Acknowledgments

This work was supported by the National Institutes of Health (NIH) under Other Transactions Authority award 1OT2DB000009-01. We are grateful to the RADx Data Hub collaborators, including RADx study investigators and the Coordination and Data Collection Centers, for their essential contributions. We also appreciate the support of the dbGaP, CEDAR, and BioPortal teams for facilitating data governance, metadata, and ontology management activities. Special thanks to the participants of RADx-funded studies for their invaluable contributions. We acknowledge the members of the RADx Data Hub User Advisory Board (UAB) for their valuable insights and contributions: Liesl Jeffers-Francis, Oswaldo Alonso Lozoya, Nader Mehri, Thais Rivas, and Peter Rose. Additionally, we thank Lucy L. Hsu, Vivian Ota Wang, and Susan K. Gregurick for their support. We are especially grateful to Mimi Duong-Quang, Sarah Lai, Tracy Maruo-Hanamoto, Ann Nguyen, Karen Stass, and Debra Vavreck, whose administrative and financial efforts were instrumental in ensuring the smooth operation of the project. We also acknowledge Josef Hardi and Rafael Gonçalves for their valuable contributions to technical discussions. Finally, we recognize the former members of the RADx Data Hub team, whose dedication and efforts have been vital to the project's success: John Graybeal, Marian Mersmann, Alexandra Shlionskaya, and Stanley C. Ahalt.

## Author Contributions

M.M.-R., J.K.Y., M.J.O., and M.U.A. led the conceptualization, drafting, and revision of the manuscript. The remaining authors reviewed the manuscript, provided critical feedback, and contributed to revisions to enhance its clarity and rigor. M.A.M., M.A.K., and A.K. supervised the project and provided strategic guidance. All authors reviewed and approved the final manuscript.

The RADx Data Hub team represents the broader consortium of official project members who have actively participated in designing, implementing, maintaining, or promoting the RADx Data Hub platform. The individual members of the RADx Data Hub team are listed below, grouped by institution and listed alphabetically by surname.

**Stanford Center for Biomedical Informatics Research, Stanford University, Palo Alto, CA 94304, USA**

Mete U. Akdogan, Luna Baalbaki, Yan Cao, Michael Dorf, Attila L. Egyedi, Crystal Han, Matthew Horridge, Christian Kindermann, Marcos Martinez-Romero, Mark A. Musen, Martin O'Connor, Alex Skrenchuk, Sowmya Somasundaram, Jimmy K. Yu

**Booz Allen Hamilton Inc., McLean, VA 22102, USA**

Mahesh Ahilandeswaran, Haven Barnes, Pierce Beckett, Johanna Brown, Beth DiGiulian, Alissa Fujimoto, Irina Gorelik, Lea Jih-Vieira, Krishnaveni Kaladi, Michael A. Keller, Marquea King, Nino Koberidze, Chun Yee Lau, Patrick Lenhart, Li Li, Amy Lin, Anna Lu, Emma Luker, Nilesh Mistry, Ashley Pryor, Bala Ramani, Joseph Schweikert, Ashley Sier, Vivian Tran, Somu Valliappan

**Renaissance Computing Institute (RENCI), University of North Carolina at Chapel Hill, Chapel Hill, NC 27517, USA**

Asiyah Ahmad, Ashley Evans, Brandy Farlow, Vicki Gardner, Suparna Goswami, Aria Henry, Robert Hubal, Yaphet Kebede, Ashok Krishnamurthy, Leah Mason, Joanna O. Mieczkowska, Keriayn Smith, Jasmine Snipe, Stephanie Suber, Josh Terrell, Aubrie Weyhmiller

## Conflicts of Interest

None declared.

## Abbreviations

NIH: National Institutes of Health
RADx: Rapid Acceleration of Diagnostics
FAIR: Findable, Accessible, Interoperable, Reusable
(C)DCC: Coordination and Data Collection Center
RADx-UP: RADx Underserved Populations
RADx-Tech: RADx Technology
RADx-rad: RADx Radical
RADx-DHT: RADx Digital Health Technologies
dbGaP: Database of Genotypes and Phenotypes
RAPIDS: Repository for RADx Digital Health Technologies
CSV: Comma-Separated Values
CEDAR: Center for Expanded Data Annotation and Retrieval
BioPortal: Ontology repository (mentioned alongside CEDAR)
PHI: Protected Health Information
PII: Personally Identifiable Information
MeSH: Medical Subject Headings
DAC: Data Access Committee
AWS: Amazon Web Services
S3: Amazon Simple Storage Service
RBAC: Role-Based Access Control
JSON-LD: JavaScript Object Notation for Linked Data
DOI: Digital Object Identifier
Jupyter: Interactive computing platform
SAS: Statistical Analysis System
ORCID: Open Researcher and Contributor Identifier
RoR: Research Organization Registry
HIPAA: Health Insurance Portability and Accountability Act
NLP: Natural Language Processing
RAS: Researcher Auth Service

## Data Availability

All study metadata and documentation are openly accessible through the RADx Data Hub Study Explorer (https://radxdatahub.nih.gov/studyExplorer). Access to study data is governed by the database of Genotypes and Phenotypes (dbGaP). Researchers can request access to study data by following the instructions available on the RADx Data Hub Resource Center (https://radxdatahub.nih.gov/resourceCenter).

## Code Availability

The software described in this paper is freely available for use at https://radxdatahub.nih.gov. The source code is openly available under the BSD 2-Clause License at https://github.com/radxdatahub.

## References

1. Kaiser J. How Much Is COVID-19 Spreading? Data Gaps and Flawed Studies Complicate Efforts to Know. *Science*. 2020;369(6500):1167-1168.

2. Mcgrail S, Dai C, Mcandrews J. Building a Data-Driven Pandemic Response: Lessons Learned from COVID-19. *Journal of Public Health Management and Practice*. 2021;27(S1):S73-S80.

3. Galaitsi SE, Cegan JC, Volk K, Joyner M, Trump BD, Linkov I. The Challenges of Data Usage for the United States' COVID-19 Response. *Int J Inf Manage*. 2021;59(102352):102352. doi:10.1016/j.ijinfomgt.2021.102352

4. Tromberg BJ, Schwetz TA, Pérez-Stable EJ, et al. Rapid Scaling Up of COVID-19 Diagnostic Testing in the United States - The NIH RADx Initiative. *N Engl J Med*. 2020;383(11):1071-1077. doi:10.1056/NEJMsr2022263

5. NIH RADx Data Hub. Accessed May 10, 2024. https://radxdatahub.nih.gov/

6. Wilkinson MD, Dumontier M, Aalbersberg IJJ, et al. The FAIR Guiding Principles for Scientific Data Management and Stewardship. *Sci Data*. 2016;3:160018. doi:10.1038/sdata.2016.18

7. RADx Programs. National Institutes of Health (NIH). June 23, 2020. Accessed November 26, 2024. https://www.nih.gov/research-training/medical-research-initiatives/radx/radx-programs

8. The RADx Data Hub Partners (DHP). RADx Data Dictionary Specification. Accessed November 29, 2024. https://github.com/bmir-radx/radx-data-dictionary-specification/blob/main/radx-data-dictionary-specification.md

9. Musen MA, Bean CA, Cheung KH, et al. The Center for Expanded Data Annotation and Retrieval. *J Am Med Inform Assoc*. 2015;22(6):1148-1152. doi:10.1093/jamia/ocv048

10. Musen MA, O'Connor MJ, Schultes E, Martínez-Romero M, Hardi J, Graybeal J. Modeling community standards for metadata as templates makes data FAIR. *Sci Data*. 2022;9(1):696. doi:10.1038/s41597-022-01815-3

11. Noy NF, Shah NH, Whetzel PL, et al. BioPortal: Ontologies and Integrated Data Resources at the Click of a Mouse. *Nucleic Acids Res*. 2009;37(Web Server issue):W170-W173. doi:10.1093/nar/gkp440

12. National Center for Biotechnology Information (NCBI). The Database of Genotypes and Phenotypes (dbGaP). Accessed November 28, 2024. https://www.ncbi.nlm.nih.gov/gap/

13. *CEDAR Embeddable Editor*. Accessed December 3, 2024.

https://github.com/metadatacenter/cedar-embeddable-editor

14. Whetzel PL, Noy NF, Shah NH, et al. BioPortal: Enhanced Functionality via New Web Services from the National Center for Biomedical Ontology to Access and Use Ontologies in Software Applications. *Nucleic Acids Res*. 2011;39(Web Server issue):W541-W545. doi:10.1093/nar/gkr469

15. RAPIDS. Accessed November 28, 2024. https://rapids.ll.mit.edu/

16. The RADx Data Hub Partners (DHP). RADx Data Hub User Code of Conduct. Accessed November 30, 2024. https://radxdatahub.nih.gov/radx-s3-resources/RADxDataHubCodeOfConduct.pdf

17. The RADx Data Hub Partners (DHP). NIH RADx Data Hub - Resource Center. Accessed November 29, 2024. https://radxdatahub.nih.gov/resourceCenter

18. Huser V, Amos L. Analyzing Real-World Use of Research Common Data Elements. *AMIA Annu Symp Proc*. 2018;2018:602-608. https://www.ncbi.nlm.nih.gov/pubmed/30815101

19. The RADx Data Hub Partners (DHP). RADx Data Hub Global Codebook. Accessed November 30, 2024. https://radxdatahub.nih.gov/radx-s3-resources/RADx_Data_Hub-Global_Codebook.xlsx

20. O'Connor MJ, Martínez-Romero M, Egyedi AL, Willrett D, Graybeal J, Musen MA. An Open Repository Model for Acquiring Knowledge About Scientific Experiments. In: *Lecture Notes in Computer Science*. Lecture notes in computer science. Springer International Publishing; 2016:762-777. doi:10.1007/978-3-319-49004-5_49

21. Section508.gov. Accessed December 2, 2024. https://www.section508.gov/

22. Cheng C, Messerschmidt L, Bravo I, et al. A General Primer for Data Harmonization. *Sci Data*. 2024;11(1):152. doi:10.1038/s41597-024-02956-3

23. Fortier I, Raina P, Van den Heuvel ER, et al. Maelstrom Research Guidelines for Rigorous Retrospective Data Harmonization. *Int J Epidemiol*. 2017;46(1):103-105. doi:10.1093/ije/dyw075

24. Cheng C, Messerschmidt L, Bravo I, et al. Harmonizing Government Responses to the COVID-19 Pandemic. *Sci Data*. 2024;11(1):204. doi:10.1038/s41597-023-02881-x

25. The RADx Data Hub Partners (DHP). RADx Data Hub Study Metadata Template. doi:10.5281/zenodo.14630796

26. RADx Data Hub Data File Metadata Template. The RADx Data Hub Partners (DHP). doi:10.5281/zenodo.14630747

27. DataCite Metadata Schema Documentation for the Publication and Citation of Research Data and Other Research Outputs — DataCite Metadata Schema 4.5 documentation. Accessed December 2, 2024. https://datacite-metadata-schema.readthedocs.io/

28. The RADx Data Hub Partners (DHP). *RADx Data Hub - De-Identification Guidance*. RADx Data Hub; 2024. Accessed November 5, 2024. https://radxdatahub.nih.gov/radx-s3-resources/RADx_Data_Hub-DeIdentification_Guidance.pdf

29. Happel C, Peñalber-Johnstone C, Tagle DA. Pivoting Novel Exosome-Based Technologies for the Detection of SARS-CoV-2. *Viruses*. 2022;14(5):1083. doi:10.3390/v14051083

30. Corbie G, D'Agostino EM, Knox S, et al. RADx-UP Coordination and Data Collection: An Infrastructure for COVID-19 Testing Disparities Research. *Am J Public Health*. 2022;112(S9):S858-S863. doi:10.2105/AJPH.2022.306953

31. Mannino RG, Nehl EJ, Farmer S, et al. The Critical Role of Engineering in the Rapid Development of COVID-19 Diagnostics: Lessons from the RADx Tech Test Verification Core. *Sci Adv*. 2023;9(14):eade4962. doi:10.1126/sciadv.ade4962

32. Waldrop AM, Cheadle JB, Bradford K, et al. Dug: A Semantic Search Engine Leveraging Peer-Reviewed Knowledge to Query Biomedical Data Repositories. *Bioinformatics*. 2022;38(12):3252-3258. doi:10.1093/bioinformatics/btac284

33. Open Researcher and Contributor Identifier (ORCID). Open Researcher and Contributor Identifier (ORCID). Accessed January 6, 2025. https://orcid.org

34. Research Organization Registry (ROR). Research Organization Registry (ROR). Accessed January 6, 2025. https://ror.org